\documentclass[aps,prl,twocolumn,10pt,notitlepage]{revtex4-1}
\usepackage{amsmath}
\usepackage{braket}
\usepackage{color}
\usepackage{graphicx}
\usepackage{subcaption}
\usepackage{caption}
\usepackage{amssymb}
\usepackage{amsfonts}
\usepackage[parfill]{parskip}
\usepackage{hyperref}
\usepackage{multirow}
\usepackage{array}
\newcommand{\swbar}[0]{\ket{s_{\bar{w}}}}
\newcommand{\ck}[0]{c^{(0)}_k}
\newcommand{\cx}[0]{c^{(0)}_x}
\newcommand{\cy}[0]{c^{(0)}_y}
\newcommand{\cl}[0]{c^{*(0)}_l}
\newcommand{\re}[0]{\text{Re}}
\begin{document}
  \title{Optimal quantum spatial search on random temporal networks}
  \date{November 29, 2017}
\author{Shantanav Chakraborty}
\author{Leonardo Novo}
\author{Serena Di Giorgio}
%\email[Correspondance:]{shantanav.chakraborty@tecnico.ulisboa.pt}
\author{Yasser Omar}
%\email[]{Your e-mail address}
%\homepage[]{Your web page}
%\thanks{}
\affiliation{Instituto de Telecomunica\c{c}\~oes, Physics of Information and Quantum Technologies Group, Lisbon, Portugal}
\affiliation{Instituto Superior T\'{e}cnico, Universidade de Lisboa, Portugal}
%%\email[]{Your e-mail address}
%%\homepage[]{Your web page}
%%\thanks{}

\begin{abstract}

To investigate the performance of quantum information tasks on networks whose topology changes in time, we study the spatial search algorithm by continuous time quantum walk to find a marked node on a random temporal network. We consider a network of $n$ nodes constituted by a time-ordered sequence of Erd\"os-R\'enyi random graphs $G(n,p)$, where $p$ is the probability that any two given nodes are connected: after every time interval $\tau$, a new graph $G(n,p)$ replaces the previous one. We prove analytically that for any given $p$, there is always a range of values of $\tau$ for which the running time of the algorithm is optimal, i.e.\ $\mathcal{O}(\sqrt{n})$, even when search on the individual static graphs constituting the temporal network is sub-optimal. On the other hand, there are regimes of $\tau$ where the algorithm is sub-optimal even when each of the underlying static graphs are sufficiently connected to perform optimal search on them. From this first study of quantum spatial search on a time-dependent network, it emerges that the non-trivial interplay between temporality and connectivity is key to the algorithmic performance. Moreover, our work can be extended to establish high-fidelity qubit transfer between any two nodes of the network. Overall, our findings show that one can exploit temporality to achieve optimal quantum information tasks on dynamical random networks.
\end{abstract}

\maketitle

Temporal networks are ubiquitous: natural, technological and social networks typically have time-varying topologies. Recently, such networks have been extensively studied at the classical level  \cite{kempe2000connectivity,mucha2010community, holme2012temporal, starnini2012random,  barabasi2016temporal}. However, quantum dynamics on temporal networks has largely been unexplored. 
Intuitively, one could expect that the uncontrolled dynamical loss and emergence of links would hinder the performance of quantum information tasks realised on networks, namely for communication, computation and sensing. But could this temporal character actually yield any advantages for such tasks? In this work, we consider the spatial search algorithm by continuous time quantum walk (CTQW) \cite{Childs_spatial_search} to find a marked node on a temporal network, and establish analytically that there are regimes where its performance is optimal.

This algorithm was first introduced in Ref.~\cite{Childs_spatial_search} and has been extensively studied on particular static graphs \cite{Meyer_symmetry,meyer2,dimred,li2016continuous}. Furthermore the analog analogue of Grover's algorithm \citep{Farhi_analog_grover} can be perceived as spatial search by quantum walk on the complete graph \cite{Childs_spatial_search}. Recently, the algorithm was proven to be optimal for Erd\"os-R\'enyi random graphs, i.e.\ graphs of $n$ nodes with each link existing between any two nodes with probability $p$  \cite{ER59, ER60}, as long as $p\geq p_{static}= \log^{3/2}(n)/n$ \cite{almost_all}. Moreover, as a random graph
can also be obtained by randomly removing links from a complete graph, these results can be seen as an analysis of the robustness of quantum search on the complete graph to random loss of links. Note that quantum dynamics on static Erd\"os-R\'enyi random graphs and other complex networks have been studied in Refs.~\cite{QCN2,QCN1}. Also, some properties of the evolution of quantum walks on dynamical percolation graphs such as the mixing time, return probabilities and spreading were studied in Refs.~\cite{leung2010coined,kollar2012coined,darazs2013continuous}.

In this paper, we study how the quantum spatial search algorithm performs on random temporal networks. These networks are obtained as a sequence of Erd\"os-R\'enyi random graphs $G(n,p)$: after every time interval $\tau$, a new graph $G(n,p)$ replaces the previous one. This problem can also
be viewed as spatial search on a complete graph with dynamical structural defects, i.e.\ where links can randomly vanish and reappear over time, as in a dynamical percolation problem.

We define the \textit{temporality} of a network as the frequency with which a given network changes its topology as compared to the relevant energy scale of the Hamiltonian representing the network, and thus  $1/\tau$ is a measure of temporality. Naturally, the introduction of this new feature leads to a much richer behaviour in the algorithmic dynamics, as compared to the static scenario. In fact, now the optimality of the algorithm depends crucially on the interplay between $\tau$ and $p$.

In our work, we find a new threshold of $p$, namely $p_{temp}=\log(n)/\sqrt{n}$, such that for $p\geq p_{temp}$ the algorithm is optimal irrespective of the temporality of the network. On the other hand, we show that sufficiently high temporality ensures that the algorithm retains its optimality for arbitrarily low values of $p$. This holds even when the underlying random graphs are comprised of mostly isolated nodes and small trees which are graphs where, in the static case, quantum search would not provide any speed-up.
\\
Interestingly there also exists an intermediate regime $p_{static}\leq p< p_{temp}$ where the spatial search algorithm is optimal on the underlying random graphs, whereas for a certain interval of $\tau$, this is no longer the case. We find that when the temporality of the network coincides with the energy scale of the Hamiltonian representing the network, the algorithmic running time is peaked. By gradually lowering or increasing the temporality, the running time of the algorithm decreases, and after a certain threshold of temporality, becomes optimal -- a behaviour also observed in Ref.~\cite{cerfroland} for the analog version of Grover's algorithm albeit in a different context. Our results show that quantum information processing tasks can be performed optimally on dynamically disordered structures.
\\
%%%%%%%%%%%%%%%%%%%%% End of Introduction %%%%%%%%%%%%%%%%%%%%%%%%
\textit{Quantum spatial search on random temporal networks.---} A temporal network is a dynamically evolving network of $n$ vertices that alters its topology after a given time interval. As a result links appear and disappear after every time interval. If initially the network is represented by a graph $G_1$, then after a time interval $\tau$, the topology of the network changes and we obtain a new graph $G_2$, and so on. Thus, within a time $t$, a temporal network may be represented by a sequence of static graphs $G_{temp}=\{G_1,G_2,...,G_m\}$, where $t=m\tau$ and $m\in \mathbb{N}$. 
\\
Naturally, a random temporal network is represented by a network that is a sequence of random graphs. Let us consider Erd\"os-R\'enyi random graphs $G(n,p)$. A random temporal network $G_{temp}(n,p,\tau)$ is a temporal sequence of Erd\"os-R\'enyi random graphs such that after a time $t=m\tau$, the network will be defined as $G_{temp}(n,p,\tau)=\{G_{1}(n,p),G_{2}(n,p),...,G_{m}(n,p)\}$, where $G_j(n,p)$ represents the random graph at the $j$th time interval. We shall focus on the optimality of the spatial search algorithm by CTQW on these networks and thus first introduce the algorithm briefly.
\\
Let $G$ represent a graph of $n$ vertices $V=\{1,...,n\}$. We consider the Hilbert space spanned by the localized quantum states at the vertices of the graph $\mathcal{H}=\text{span}\{\ket{1},\ket{2},...,\ket{n}\}$. The search Hamiltonian corresponding to $G$ is given by
\begin{equation}
\label{eq:search_ham}
H_{search}=-E\ket{w}\bra{w}-\gamma A_G,
\end{equation}
where $\ket{w}$ corresponds to the solution node of the search problem marked by the local site energy $E$, $\gamma$ is a real number and $A_G$ is the adjacency matrix of the graph $G$ \footnote{The search Hamiltonians $H_{search}=\ket{w}\bra{w}+\gamma A_G$ and $H_{search}=-\ket{w}\bra{w}-\gamma A_G$ are equivalent and leave the outcome of the algorithm unaltered. We shall use them interchangeably.}. We set the energy scale $E$ to be $1$, such that the quantum simulation of $\ket{w}\bra{w}$ for time $t$ would correspond to $\mathcal{O}(t)$ queries to the standard Grover oracle \citep{Childs_spatial_search}. The initial state of the algorithm is usually chosen to be the equal superposition of all vertices, i.e.\ the state $\ket{s}=\sum_{i=1}^n \ket{i}/\sqrt{n}$. The quantum search algorithm is said to be optimal on graph $G$ if there exists a value of $\gamma$ such that after a time $T=\mathcal{O}(\sqrt{n})$, the probability of obtaining the solution upon a measurement in the basis of the vertices is $|\braket{w|e^{-iH_{search}T}|s}|^2=\mathcal{O}(1)$ \cite{Childs_spatial_search}.
\\
In order to analyze this algorithm on $G_{temp}(n,p,\tau)$, we use two separate approaches to prove our results for different ranges of $p$. For $p\geq p_{static}=\log^{3/2}(n)/n$, we use the fact that the maximum eigenvalue of the adjacency matrix of each of the random graphs appearing during the time evolution is separated from the bulk of the spectrum, and the eigenstate corresponding to it is almost surely the initial condition of the algorithm $\ket{s}$, as was shown to be the case in Lemma 2 of Ref.~\cite{almost_all}. To obtain the regime where the optimality of the algorithm is maintained as a function of $\tau$ and $p$,  we use time-dependent perturbation theory. However, this property about the spectrum of adjacency matrices of random graphs does not hold when $p$ is below the aforementioned threshold. So, for such regimes, we construct a linear superoperator that describes the average dynamics of the algorithm on random temporal networks. We present each of these approaches separately.\\
\textit{Quantum spatial search on random temporal networks having $p\geq p_{static}$.---} As long as $p\geq\log^{3/2}(n)/n$, the eigenstate corresponding to the maximum eigenvalue of the adjacency matrix of an Erd\"os-R\'enyi random graph is almost surely the state $\ket{s}$ with eigenvalue $np$ \cite{almost_all}. Thus the adjacency matrix of each of the random graphs appearing in $G_{temp}(n,p,\tau)$ satisfies this property. Let $A_j$ denote the adjacency matrix of the random graph appearing at the $j$th time instance (i.e.\ after a time $t=j\tau$). Then, each off-diagonal entry of $A_j$ is $1$ with probability $p$ and $0$ with probability $1-p$. Let $B_j=A_j-np\ket{s}\bra{s}+pI$ where $B_j$ is a random matrix with each off-diagonal entry having mean $0$ and variance $p$, with the diagonal entries being zero, and $I$ is the identity matrix. We define the search Hamiltonian for $G_{temp}(n,p,\tau)$ as in Eq.~\eqref{eq:search_ham} by choosing $\gamma=1/(np)$.  By expressing each of the adjacency matrices appearing in $G_{temp}(n,p,\tau)$ as mentioned previously, we obtain the following search Hamiltonian: 
\begin{equation}
\label{eq:search_ham_temporal}
H_{search}(t)=\underbrace{-\ket{w}\bra{w}-\ket{s}\bra{s}}_{H_0}-\underbrace{\sum_{j=1}^{m}\gamma B_jf_j(t,\tau)}_{V(t)},
\end{equation}
where $f_j(t,\tau)=\Theta(t-(j-1)\tau)-\Theta(t-j\tau)$, where $\Theta(x)$ is the Heaviside function, and $m=T/\tau$ is the number of instances of random graphs appearing throughout the evolution time of $T=\mathcal{O}(\sqrt{n})$. Here $H_0$ induces a rotation in the two dimensional subspace spanned by $\ket{w}$ and $\ket{s}$, whereas $V(t)$ will induce a coupling between this subspace and the $n-2$ degenerate eigenspace of $H_0$.  
\begin{figure}[h!]
\centering
\includegraphics[scale=0.5]{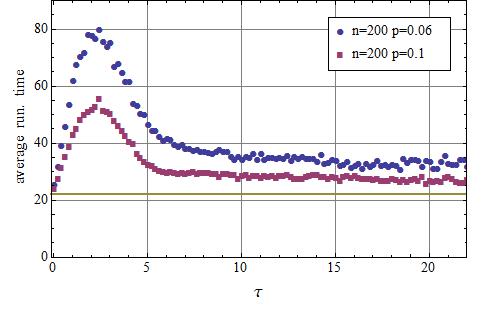}
\vspace{-22 pt}
\caption{\small{Average running time of the quantum spatial search algorithm as a function of $\tau$ for $G_{temp}(200,0.06,\tau)$ (in blue dots) and $G_{temp}(200,0.1,\tau)$(in red squares). Each point is averaged over 100 realizations. As predicted, the average running time peaks at $\tau\sim 1$, when the temporality coincides with the energy scale of the search Hamiltonian. Away from this peak, the average running time decreases gradually towards the optimal running time (indicated by the solid line).}}
\label{fig_temporal1}
\end{figure}
Also, $H_0$ is the search Hamiltonian corresponding to the quantum walk on a complete graph, where the search algorithm runs optimally \cite{Farhi_analog_grover, Childs_spatial_search}. In this case, we treat $V(t)$ as a perturbation to $H_0$ and use time-dependent perturbation theory. Let $\ket{\psi(t)}$ be the wavefunction of the quantum walk  obtained by evolving under $H_{search}(t)$. The error probability induced by the perturbation is thus $\epsilon=1-|\braket{w|\psi(T)}|^2$, where $T=\mathcal{O}(\sqrt{n})$.
\\
We are interested in calculating when the average error probability $\langle\epsilon\rangle$ is bounded for a given $\tau$ and $p$. Whenever $\langle\epsilon\rangle\sim o(1)$, the algorithm outputs the solution state $\ket{w}$ with probability $1-o(1)$ in $\mathcal{O}(\sqrt{n})$ time. Without loss of generality, we intend to bound $\langle\epsilon\rangle=\mathcal{O}(1/\log(n))$ \footnote{In our analysis when $p\geq\log^{3/2}(n)/n$, we neglect terms of $\mathcal{O}(1/(np))$ and we fix $\braket{\epsilon}=1/\log(n)$ to be the upper bound of error throughout and obtain bounds for $\tau$ and $p$.}. We prove that the average error probability is given by (for the derivation see Sec.~I of the Supplemental Material):
\begin{equation}
\label{eq:average_error_main}
\langle\epsilon\rangle = 
  \begin{cases} 
   \mathcal{O}\left(\dfrac{\tau}{p\sqrt{n}}\right) & \text{if } \tau <\mathcal{O}(1) \\
      \mathcal{O}\left(\dfrac{1}{p\tau\sqrt{n}}\right)    & \text{if } \tau \geq \mathcal{O}(1)
  \end{cases}.
\end{equation} 
Firstly, we are interested in finding the regime of $p$ for which the algorithm is robust to temporality. From Eq.~\eqref{eq:average_error_main} we find that as long as $p\geq p_{temp}=\log(n)/\sqrt{n}$, the average error is bounded irrespective of any $0<\tau\leq \mathcal{O}(\sqrt{n})$. For lower values of $p$, temporality becomes crucial to the optimality of the algorithm and in fact for the range of $p$ between $p_{static}$ and $p_{temp}$ there exist two separate regimes of temporality that determine the optimality of the algorithm: a fast temporality regime and a slow temporality regime such that if the topology of the network alters faster than $\tau_{\text{fast}}=\mathcal{O}(p\sqrt{n}/\log(n))$ or slower than $\tau_{\text{slow}}=\mathcal{O}(\log(n)/(p\sqrt{n}))$, the algorithm remains optimal. The behavior of the algorithm in the intermediate regime of $\tau_{\text{fast}}<\tau<\tau_{\text{slow}}$ is also interesting, albeit suboptimal. As the temporality of the network increases from $\tau_{\text{fast}}$, the algorithmic running time increases with it peaking at $\tau=\mathcal{O}(1)$, after which it gradually decreases  until $\tau=\tau_{\text{slow}}$. To confirm this, we plot in Fig.~\ref{fig_temporal1} the average running time of $G_{temp}(200,0.06,\tau)$ (blue dots) and $G_{temp}(200, 0.1,\tau)$ (red squares) as a function of $\tau$. As predicted, the average running time peaks when the temporality $1/\tau\approx 1$ and approaches the optimal running time (solid line) away from the peak.
\\
A similar behaviour has also been observed in Ref.~\cite{cerfroland} for the analog version of Grover's algorithm, for the following noise model: the authors consider a perturbation to the search Hamiltonian in the form of a random matrix, with each entry being a time dependent random variable with a predefined autocorrelation function and with a certain cut-off frequency. The authors find that when the cut-off frequency of noise scales much faster or slower than the energy scale of the Hamiltonian, the algorithm retains its optimality. On the other hand, when they scale similarly (i.e.\ when the cut-off frequency of noise is $\mathcal{O}(1)$), the average error is bounded by a constant only when the ratio of the norm of the perturbation Hamiltonian and that of the unperturbed search Hamiltonian scales as $\mathcal{O}(n^{-1/4})$. Analogously, we find that for networks with constant temporality, the average error is constant when $p\sim 1/\sqrt{n}$, in which case the aforementioned ratio is  also $||V(t)||/||H_0||=||B_j||=\mathcal{O}(n^{-1/4})$, where we have used the fact that $||B_j||=\mathcal{O}(\sqrt{np})$ \cite{furedi1981eigenvalues, vu2007}. This shows that the global features of the response of this algorithm with respect to the typical noise time scales for these two models is quite similar. 
\\
Note that we also recover the scenario of the spatial search algorithm on a static random network by choosing $\tau=\mathcal{O}(\sqrt{n})$. In this case the average error is always bounded for $p\geq p_{static}$, thereby recovering the results of Ref.~\cite{almost_all}.
\\
%%%%%%%%%%%%%%%%%%%%%%%%%%%%%%%%%%%%%%%%%%%%%%%%%%%%%%%%%%%%%
\textit{Quantum spatial search on random temporal networks having $p< p_{static}$.---} 
~Here we prove that for random temporal networks with sufficiently high temporality, the spatial search algorithm is optimal for arbitrarily low $p$. For this regime of $p$, the results obtained previously no longer hold, as $\ket{s}$ is not an eigenstate of the adjacency matrix (and $np$ is no longer the maximum eigenvalue) of an Erd\"os-R\'enyi random graph. For $p<\log(n)/n$, the underlying random graphs are no longer connected \cite{barabasi_review}. Moreover, for $p<<1/n$, the static random graphs appearing during the time evolution of the algorithm are extremely sparse and are mostly comprised of isolated nodes and trees. In particular, we shall focus on finding a regime of optimality of the search algorithm for $p\leq 1/n$, while we refer the readers to Sec.~II of the Supplemental Material for results when $1/n<p<p_{static}$.
\\
In this regime we follow a different approach: we consider the evolution of the quantum state averaged over all possible realizations of a random graph using the density matrix formalism. The number of possible realizations of $G(n,p)$ is $|G|=2^{N}$, where $N=\binom{n}{2}$. The average dynamics of the algorithm on a random temporal network after one time step $\tau$ is described by the following superoperator:
\begin{align}
\label{eq:superoperator_definition_main}
\Phi(\rho)=\sum_{r=1}^{|G|}p_r e^{-iH_r\tau}\rho e^{iH_r\tau}=\langle e^{-iH_r\tau}\rho e^{iH_r\tau}\rangle,
\end{align}
where $p_r$ is the probability of the $r$th realization and $H_r=\ket{w}\bra{w}+\gamma A_{G_r}$ with $A_{G_r}$ being the adjacency matrix corresponding to the $r$th realization of $G(n,p)$. Let $\langle X \rangle$ represent the expected value of $X$. The evolution of the algorithm after $m=\mathcal{O}(\sqrt{n}/\tau)$ time steps is given by $\Phi^m(\rho)$. The first order expansion of the superoperator yields:
\begin{align}
\Phi(\rho)&=\left\langle\rho-i\tau[H_r,\rho]\right\rangle+\delta\\
          &=\rho-i\tau[\ket{w}\bra{w}+\ket{s}\bra{s},\rho]+\delta\\
          &=\Phi_0(\rho)+\delta,    
\end{align}
where the second step follows because the expected value of each entry of $A_{G_r}$ is $p$ and so $\langle A_{G_r}\rangle=np\ket{s}\bra{s}$. Thus $\langle H_r\rangle=\ket{w}\bra{w}+\ket{s}\bra{s}$, which is the same as $H_0$ defined in Eq.~\eqref{eq:search_ham_temporal}, and is optimal for quantum spatial search. Here $\delta$ is the error induced by truncating the superoperator $\Phi$ after the first order and is given by $\delta\leq\sum_{k=2}^{\infty}(\tau^k/k!)\left\langle||H^k_r||\right\rangle$. 
\begin{figure}[h!]
\centering
\includegraphics[scale=0.55]{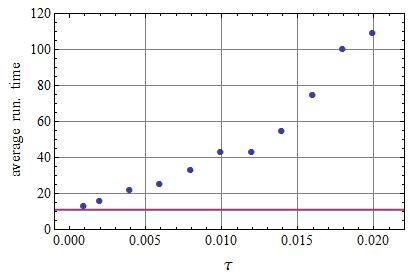}
\vspace{-12pt}
\caption{\small{Average running time of the quantum spatial search algorithm on $G_{temp}(50,0.0008,\tau)$ (dots) as a function of $\tau$. Each point is averaged over 50 realizations. Note that $p=2/n^2$ and even then, for small enough $\tau$, the algorithm runs in optimal time (solid line). As $\tau$ is increased, the algorithmic running time increases.}}
\label{fig_temporal2}
\end{figure}
Note that the superoperator $\Phi_0^m$ describes approximately the standard evolution of the algorithm under the Hamiltonian $\langle H_r\rangle$, and thus we intend to bound the error obtained by using the superoperator $\Phi$ instead of the superoperator $\Phi_{0}$ to describe the dynamics for each of the $m$ timesteps. This is given by :
\begin{equation}
\label{eq:error_belowperc_main}
\epsilon=||\Phi^m(\rho)-\Phi^m_0(\rho)||\leq m\delta.
\end{equation}
Thus to bound $\epsilon$ we need to bound $\langle ||A_{G_r}||\rangle$. Since $p\leq 1/n$, the underlying random networks are extremely sparse, containing isolated nodes and few links. Thus $||A_{G_r}||$ is bounded by the sum of the individual links of the random graphs. As $p$ decreases further (i.e. $p<<1/n$), the aforementioned bound is better as the underlying networks have fewer and fewer links. For a given range of $p$ we find the bound for $\tau$ where $\epsilon\leq\mathcal{O}(1/\log(n))$. In fact we obtain that (for the derivations refer to Sec.~II of Supplemental Material)
\begin{equation}
\label{eq:average_error_belowperc_main}
\tau \leq 
  \begin{cases} 
   \dfrac{1}{n^{5/2}\log(n)} & \text{if } 1/n^2 \leq p\leq 1/n \\
      \dfrac{p}{\sqrt{n}\log(n)}    & \text{if } p<1/n^2
  \end{cases}.
\end{equation} 
In general our results imply that although $p$ is well below the percolation threshold, and in fact the temporal network consists of graphs that do not have giant components and are mostly composed of isolated nodes and trees of $\mathcal{O}(1)$ nodes, sufficiently high temporality can still lead to optimal search. This cannot be achieved by performing a quantum walk on any of these structures appearing as a static network.
\begin{figure*}
  \includegraphics[width=\textwidth,height=2.5cm]{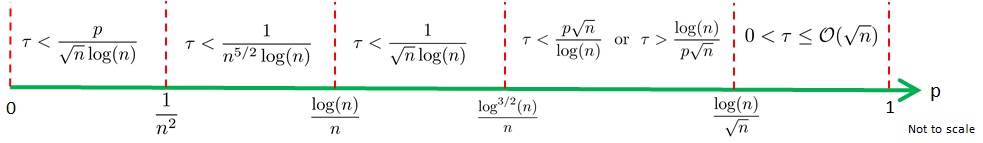}
  \vspace{-12pt}
\caption{\small{Summary of analytical results: thresholds of $\tau$ above or below which the quantum spatial search algorithm on a random temporal network of $n$ nodes is optimal for a given range of $p$.}}
\label{fig_temporal3}
\end{figure*}
This has been confirmed in Fig.~\ref{fig_temporal2}, wherein we plot (in blue points) the average running time of the quantum spatial search algorithm on random temporal networks $G_{temp}(50,0.0008,\tau)$ with a value of $p$ is way below percolation threshold $(p=2/n^2)$. As expected, for sufficiently low values of $\tau$, the running time of the algorithm is close to the optimal running time of $T=\pi\sqrt{n}/2$ (solid line) and increases as $\tau$ is increased. We summarize the regimes of $\tau$ and $p$ where the algorithm is optimal in Fig.~\ref{fig_temporal3}. See Supplemental Material for derivations.
\\
%%%%%%%%%%%%%%%%%%%%%%%%%%%%%%%%%%%%%%%%%%%%%%%%%%%%%%%%%%%%%%%%%%

\textit{Discussion.---} We have proven analytically that for any given $p$, there is always a range of values of $\tau$ for which the running time of the spatial search algorithm by CTQW on a random temporal network $G_{temp}(n,p,\tau)$ is optimal, i.e.\ $\mathcal{O}(\sqrt{n})$. Indeed, we find that the non-trivial interplay between $p$ and the temporality of the network is key to the algorithm's performance (see Fig.~\ref{fig_temporal3}).

We obtain a threshold $p_{temp}=\log(n)/\sqrt{n}$ above which the algorithm is optimal irrespectively of $\tau$, i.e.\ of how fast or slowly the links appear and disappear in the dynamical network.

We also find that, for sufficiently low values of $\tau$, the algorithm is optimal for any value of $p$. This means that high temporality allows optimal performance even when $p$ is well below the static percolation threshold, i.e.\ when the underlying static graphs are comprised mostly of isolated nodes and trees of constant depth.
  
Interestingly, for $p_{static}<p<p_{temp}$, the algorithm is optimal on each static random graph, but not always on the temporal network composed by the sequence of such graphs. In the sub-optimal regime, the algorithmic running time is peaked when the temporality of the network coincides with the energy scale of the search Hamiltonian. We can move away from this regime by decreasing or increasing the temporality: the running time of the algorithm will then decrease accordingly, reaching the optimal performance at $\tau_\text{slow}$ or $\tau_\text{fast}$ respectively.

Note that our results on spatial search can also be extended to perform high-fidelity state transfer of a qubit between any two nodes of a random temporal network \cite{hein2009wave,tanner_graphene,almost_all}.

Finally, our findings can also be interpreted as an analysis of the robustness of the quantum spatial search algorithm and the state transfer protocol on a complete graph with dynamical structural defects. Furthermore, they pave the way to study quantum dynamics on non-Markovian temporal networks \cite{scholtes2014causality}, as well as to exploit temporality as a control mechanism to improve or protect the effectiveness and efficiency of quantum information tasks on dynamical networks.
\\

\begin{acknowledgments}
\textit{Acknowledgments} --- The authors thank the support from Funda\c{c}\~{a}o para a Ci\^{e}ncia e a Tecnologia (Portugal), namely through programmes PTDC/POPH/POCH and projects UID/EEA/50008/2013,~IT/QuSim,~IT/QuNet,~ProQuNet,~partially funded by EU FEDER,  from the EU FP7 project PAPETS (GA 323901), and from the JTF project \textit{The Nature of Quantum Networks} (ID 60478). Furthermore SC, LN and SDG acknowledge the support from the DP-PMI and FCT (Portugal) through SFRH/BD/52246/2013, SFRH/BD/52241/2013 and PD/BD/114332/2016 respectively.
\end{acknowledgments}
%%%%%%%%%%%%%%%%%%%%%%%%%%%%%%%%%%%%%%%%%%%%%%%%%%%%%%%%%%%%%
\widetext
%\clearpage
\begin{center}
\textbf{\large Supplemental Material}
\end{center}
\vspace{0.5cm}
\setcounter{equation}{0}
\setcounter{figure}{0}
\setcounter{table}{0}
\makeatletter
\renewcommand{\theequation}{S\arabic{equation}}
\renewcommand{\thefigure}{S\arabic{figure}}
In the main letter we have that $p_{static}=\log^{3/2}(n)/n$.
\\
\\
\textbf{I.~Optimality of quantum search on random temporal networks when $p\geq p_{static}$---} Let $A_j$ denote the adjacency matrix of the random graph appearing at the $j$th time instance (i.e.\ after a time $t=j\tau$). Then each off-diagonal entry of $A_j$ is $1$ with probability $p$ and $0$ with probability $1-p$. Let
\begin{equation}
B_j=A_j-np\ket{s}\bra{s}+pI.
\end{equation} 
The matrix $B_j$ is a random matrix with each off-diagonal entry having mean $0$ and variance $p$ with the diagonal entries being zero, and $I$ is the identity matrix. 
%As long as $p\geq\log^{3/2}(n)/n$, the eigenstate corresponding to the maximum eigenvalue of the $A_j$ is almost surely the state $\ket{s}$ with eigenvalue $np$. Thus, the adjacency matrix of each of the random graphs appearing in $G_{temp}(n,p,\tau)$ would satisfy this property. So we have that, for all $j$ 
%\begin{align}
%\label{eq:orthogonality}
%\braket{s|\gamma B_j|s}&=\frac{1}{n}\approx 0.~~~~~~\text{(We neglect terms of $\mathcal{O}(1/n)$)}
%\end{align} 
%We shall use this property in our derivation at a later stage.
We have that 
\begin{equation}
H_{search}(t)=-\ket{w}\bra{w}-\ket{s}\bra{s}-\sum_{j=1}^{m}\gamma B_jf_j(t,\tau),
\end{equation}
where $f_j(t,\tau)=\Theta(t-(j-1)\tau)-\Theta(t-j\tau)$ with $\Theta(x)$ being the Heaviside function. Let $H_0=-\ket{w}\bra{w}-\ket{s}\bra{s}$ and $V(t)=\sum_{j=1}^{m}\gamma B_jf_j(t,\tau)$. Let $\swbar$ represent the state that is an equal superposition of all nodes of the network other than the solution. The ground state and first excited state of $H_0$ are $\ket{\lambda_1}=(\ket{w}+\swbar)/\sqrt{2}$ and $\ket{\lambda_2}=(\ket{w}-\swbar)/\sqrt{2}$ respectively with the corresponding energies being $\lambda_1=-1-1/\sqrt{n}$ and $\lambda_2=-1+1/\sqrt{n}$. The remaining $n-2$ eigenstates form a degenerate subspace of energy $0$. Without loss of generality let the solution node $\ket{w}=\ket{1}$ and then the remaining eigenstates are represented as   
\begin{equation}
\label{eq:degenerate_subspace}
\ket{\lambda_k}= \frac{1}{\sqrt{n-1}}\sum_{j=2}^{n}\omega^{(k-2)j}\ket{j},~3\leq k\leq n,
\end{equation}
where $\omega=e^{i2\pi/(n-1)}$. 

We treat $V(t)$ as a perturbation to $H_0$ and use time-dependent perturbation theory. Let the wavefunction corresponding to the evolution under $H_0$ after a time $t$ be 
\begin{equation}
\label{eq:unperturbed_ham}
\ket{\phi(t)}=\sum_{k=1}^{n}\ck e^{-i\lambda_k t}\ket{\lambda_k}.
\end{equation}
As $H_0$ is optimal for quantum spatial search, after a time $T=\mathcal{O}(\sqrt{n})$, the wavefunction is localized at the solution node $\ket{w}$, i.e.
\begin{equation}
\label{eq:solution}
\ket{\phi(T)}=\ket{w}=\frac{\ket{\lambda_1}+\ket{\lambda_2}}{\sqrt{2}}.
\end{equation}
Now, in the presence of the time-dependent perturbation term $V(t)$, after the time $t$, assume that we would obtain the wavefunction $\ket{\psi(t)}$. The error probability induced by the perturbation is thus 
\begin{equation}
\epsilon=1-|\braket{w|\psi(T)}|^2.
\end{equation}
We are interested in calculating when the average error is bounded as a function of $\tau$ and $p$. 

Using time-dependent perturbation theory, we obtain that
\begin{equation}
\ket{\psi(t)}=\sum_{k=1}^{n}c_{k}(t)e^{-i\lambda_k t}\ket{\lambda_k}.
\end{equation}
Here
\begin{equation}
\label{eq:perturbation_expansion}
c_k(t)=\ck-i\sum_{x}\cx\int_{0}^{t}e^{\omega_{kx}t_1}v_{kx}(t)dt-\sum_{x,y}\cy\int_{0}^{t}\int_{0}^{t_1}e^{i(\omega_{kx}t_1+\omega_{xy}t_2)}v_{kx}(t_1)v_{xy}(t_2)dt_1dt_2,
\end{equation}
where $\omega_{kx}=\lambda_k-\lambda_x$ and 
\begin{equation}
\label{eq:overlap-eigenspace}
v_{kx}(t)=\sum_{j=1}^{m}\braket{\lambda_k|\gamma B_j|\lambda_x}f_j(t,\tau).
\end{equation}
Observe that each diagonal entry of $B_j$ is $0$ and all other entries have mean $0$. So
\begin{equation}
\label{eq:expectation_single_v}
\mathbb{E}[v_{kx}(t)]=0,
\end{equation} 
where $\mathbb{E}[X]$ represents the expected value of random variable $X$. Also from Eq.~\eqref{eq:solution} we find that the solution state $\ket{w}$ has non-zero overlap in only the ground and first excited states of $H_0$. This yields
\begin{equation}
\label{eq:overlap_with_solution}
|\braket{w|\psi(t)}|^2=\frac{1}{2}\Big{[}|c_1(t)|^2+|c_2(t)|^2\Big{]}+\re\Big{[}c_1^*(t)c_2(t) e^{-i\omega_{12}t}\Big{]}.
\end{equation}
In general in calculating Eq.~\eqref{eq:overlap_with_solution}, we obtain terms such as
\begin{align}
\begin{split}
c^{*}_{l}(t)c_{k}(t)&=\ck\cl-i\cl\sum_{y}c^{(0)}_{y}\int_{0}^{t}e^{i\omega_{ly}t_1}v_{ly}(t_1)dt_1+i\ck\sum_{y}c^{*(0)}_{y}\int_{0}^{t}e^{-i\omega_{ky}t_1}v^{*}_{ky}(t_1)dt_1\\
                    &+\sum_{x,y}c^{*(0)}_{x}c^{(0)}_{y}\left(\int_{0}^{t}e^{-i\omega_{kx}t_1}v^{*}_{kx}(t_1)dt_1\right)\left(\int_{0}^{t}e^{i\omega_{ky}t_1}v_{ky}(t_1)dt_1\right)\\
                    &-\sum_{x,y}\ck c^{*(0)}_y\int_{0}^{t}\int_{0}^{t_1}e^{-i(\omega_{lx}t_1+\omega_{xy}t_2)}v^{*}_{lx}(t_1)v^{*}_{xy}(t_2)dt_1dt_2\\
                    &-\sum_{x,y}\cl c^{(0)}_y\int_{0}^{t}\int_{0}^{t_1}e^{i(\omega_{kx}t_1+\omega_{xy}t_2)}v_{kx}(t_1)v_{xy}(t_2)dt_1dt_2.
\end{split}
\label{eq:each_term_overlap_with_solution}
\end{align}
As we are interested in calculating the average error $\braket{\epsilon}$, the first order terms of Eq.~\eqref{eq:each_term_overlap_with_solution} are going to be zero owing to Eq.\eqref{eq:expectation_single_v}. Also from Eq.~\eqref{eq:overlap_with_solution} we have that the indices $k,l\in\{1,2\}$. The initial state of the search algorithm is $\ket{s}=(\ket{\lambda_1}-\ket{\lambda_2})/\sqrt{2}$. So only the ground state and the first state are initially populated. As such $\ck=0$, for all $k>2$. So in each of the double integrals, two out of three indices must be in $\{1,2\}$ and the other index may be greater than $2$. The second order terms require calculating correlation functions of the following form
\begin{align}
\label{eq:autocorrelation}
\braket{v_{ab}(t_1)v^{*}_{ac}(t_2)}&=\sum_{x,y=1}^{m}\mathbb{E}[\braket{\lambda_a|\gamma B_x|\lambda_b}\braket{\lambda_c|\gamma B_y|\lambda_a}]f_x(t_1,\tau)f_y(t_2,\tau)\delta_{xy}.
\end{align}
To calculate these correlation functions we make use of the fact that $B_x$ is a random matrix such that each non-diagonal entry is a random variable with mean $0$ and variance $p$ and all diagonal entries being $0$.
When $a,b,c\in\{1,2\}$, we need to calculate quantities such as 
\begin{align}
\label{eq:autocorrelation_swbar}
\begin{split}
\mathbb{E}[\braket{s_{\bar{w}}|\gamma B_x|s_{\bar{w}}}\braket{s_{\bar{w}}|\gamma B_x|s_{\bar{w}}}]&=\dfrac{1}{(n-1)^2}\mathbb{E}[\sum_{p,q,r,s=2}^{n}\braket{p|\gamma B_x|q}\braket{r|\gamma B_x|s}]\\
&=\dfrac{1}{(n-1)^2}\mathbb{E}\left[\sum_{i,j=2;i\neq j}^{n}\braket{i|\gamma B_x|j}\braket{j|\gamma B_x|i}\right]\\
&=\dfrac{1}{(n-1)^2}\sum_{i,j=2;i\neq j}^{n}\mathbb{E}\left[|\braket{i|\gamma B_x|j}|^2\right]\\
&\approx\dfrac{(n-1)^2\gamma^2 p}{(n-1)^2}=\dfrac{1}{n^2p},
\end{split}
\end{align}
where we arrive at the second line from the first by considering the fact that independent entries of $B_x$ are uncorrelated. Also $\mathbb{E}[\braket{w|\gamma B_x|w}\braket{w|\gamma B_x|w}]=0=\mathbb{E}[\braket{w|\gamma B_x|s_{\bar{w}}}\braket{s_{\bar{w}}|\gamma B_x|s_{\bar{w}}}]$ and $\mathbb{E}[\braket{w|\gamma B_x|s_{\bar{w}}}\braket{s_{\bar{w}}|\gamma B_x|w}]=1/(n^2p)$. So for all $a,b,c\in\{1,2\}$ we find that $\braket{v_{ab}(t_1)v^*_{ac}(t_2)}=\mathcal{O}(1/(n^2p))$. Now assume in Eq.~\eqref{eq:autocorrelation} that $a=1$ and $b=c>2$. We have
\begin{align}
\label{eq:autocorrelation_twodimensional_degenerate}
\begin{split}
\braket{v_{1b}(t_1)v^{*}_{1c}(t_2)}&=\sum_{x=1}^{m}\mathbb{E}[\braket{\lambda_1|\gamma B_x|\lambda_b}\braket{\lambda_c|\gamma B_x|\lambda_1}]f_x(t_1,\tau)f_x(t_2,\tau)\\
                                   &=\dfrac{1}{2}\big{\{}\sum_{x=1}^{m}\underbrace{\mathbb{E}[\braket{w|\gamma B_x|\lambda_b}\braket{\lambda_c|\gamma B_x|w}]}_{\mathbb{E}^{x}_{1bc}[...]}+\mathbb{E}[\braket{s_{\bar{w}}|\gamma B_x|\lambda_b}\braket{\lambda_c|\gamma B_x|w}]+\mathbb{E}[\braket{w|\gamma B_x|\lambda_b}\braket{\lambda_c|\gamma B_x|s_{\bar{w}}}]\\
                           &\text{~~~~~~~~~~~~~~~~}+\mathbb{E}[\braket{s_{\bar{w}}|\gamma B_x|\lambda_b}\braket{\lambda_c|\gamma B_x|s_{\bar{w}}}]\big{\}}f_x(t_1,\tau)f_x(t_2,\tau).\\                                
\end{split}
\end{align}
Now using the fact that independent entries of the random matrix $B_j$ are uncorrelated, we have terms such as
\begin{align}
\mathbb{E}^{x}_{1bc}[...]&=\frac{1}{2(n-1)}\mathbb{E}[\sum_{\alpha,\beta=2}^{n}\braket{w|\gamma B_x|\alpha}\omega^{(b-2)\alpha}\braket{\beta|\gamma B_x|w}\omega^{-(c-2)\beta}\delta_{\alpha\beta}\delta_{bc}]\\ 
                        &=\frac{1}{2(n-1)}\sum_{\alpha=2}^{n}\gamma^2p=\frac{1}{2n^2p}\text{~~~~~[As for each non-diagonal element $(x,y)$ of $B_j$,~$\mathbb{E}[|\braket{x|B_j|y}|^2]=p$~]}.
\end{align}
All other terms of the sum in Eq.~\eqref{eq:autocorrelation_twodimensional_degenerate} would give the same value and so the sum would be $\chi=\mathcal{O}(1/(n^2p))$. Thus,
\begin{equation}
\label{eq:autocorrelation_final}
\braket{v_{1b}(t_1)v^{*}_{1c}(t_2)}=\chi\sum_{x=1}^{m}f_x(t_1,\tau)f_x(t_2,\tau).
\end{equation}
We find that the autocorrelation function obtained in Eq.~\eqref{eq:autocorrelation_final} has the same form even when $a,b\in \{1,2\}$ and $c>2$. From Eq.~\eqref{eq:each_term_overlap_with_solution} we find that as $\ck=0$ for $k>2$, and so the integrals where $a\in\{1,2\}$ and $b,c>2$ such that $b\neq c$ would not appear when considering second order perturbation theory. So the result in Eq.~\eqref{eq:autocorrelation_final} is independent of $b$ and $c$. For any $b,c>2$ the same result would be obtained.  Also the term where all of $a,b,c >2$ are never encountered.\\
Subsequently, we calculate the average error after a time $T=\pi\sqrt{n}/2$.
This involves calculating integrals of the following forms
\begin{align}
\label{eq:integral_1}
I^{\pm}_{xyz}&=\chi\sum_{j=1}^{m}\int_{(j-1)\tau}^{j\tau}\int_{(j-1)\tau}^{t_1}dt_1 dt_2 e^{\pm i(\omega_{xz}t_1-\omega_{yz}t_2)}.
\end{align}
When $x,y,z\in\{1,2\}$ these integrals are at most $\mathcal{O}(1/(np))$. So as $p\geq \log(n)/n$, these integrals would be upper bounded by $\mathcal{O}\left(1/\log(n)\right)$. Now there are a fixed number of such double integrals that contribute to the average error $\langle\epsilon\rangle$. Hence $\langle\epsilon\rangle\leq\mathcal{O}\left(1/\log(n)\right)$ as a consequence of these double integrals.

In fact the terms that would contribute significantly to $\langle\epsilon\rangle$ are the terms that couple the $(n-2)$ fold degenerate subspace with the ground and first excited states as there are $(n-2)$ such integrals. For this we consider integrals in Eq.~\eqref{eq:integral_1} such that where $x,y\in \{1,2\}$ and $z>2$. Also the correlation functions in $I^{\pm}_{xyz}$ are the same for any $z>2$. Thus it suffices to replace $z$ by $3$ and we have that
\begin{align}
\label{eq:integral_general_supp}
I^{\pm}_{xy3}&=
\begin{cases}
\chi\sum_{j=1}^{m}\left(\pm i\dfrac{\tau}{\omega_{x3}}
+\dfrac{1-e^{\pm i\omega_{x3}\tau}}{\omega^2_{x3}}\right), 
\text{~~if $x= y$}\\~\\
\chi\sum_{j=1}^{m}
\left(
\dfrac{e^{\pm i(\omega_{x3}-\omega_{y3})(j-1)\tau}(e^{\pm i(\omega_{x3}-\omega_{y3})\tau}-1)}{\omega_{y3}(\omega_{x3}-\omega_{y3})}
+
\dfrac{e^{\pm i(\omega_{x3}-\omega_{y3})(j-1)\tau}-e^{\pm i(\omega_{x3}j\tau-\omega_{y3}(j-1)\tau)}}{\omega_{x3}\omega_{y3}}
\right),
 \text{~~if $x\neq y$}
\end{cases}
\end{align}
The average probability of error is given by 
\begin{equation}
\label{eq:error_supp}
\braket{\epsilon}=\dfrac{n-2}{2}\left(\sum_{x,y=1}^{2}(I^{+}_{xy3}+I^{-}_{xy3})+2\re[e^{-i\omega_{12}T}(I^{-}_{113}+I^{+}_{223}+I^{-}_{123}+I^{+}_{213})]\right)+\mathcal{O}\left(\dfrac{1}{np}\right)
\end{equation}
The first set of integrals satisfy:
\begin{equation}
\label{eq:sum_of_integrals}
\sum_{x=1}^{2}(I^{+}_{xx3}+I^{-}_{xx3})=\mathcal{O}\left(\dfrac{\gamma^2 p\sqrt{n}}{\tau}\sum_{x=1}^{2}\dfrac{\sin^2(\omega_{x3}\tau/2)}{\omega_{x3}^2}\right).
\end{equation}
On the other hand, the remaining integrals satisfy:
\begin{equation}
\sum_{x,y=1;x\neq y}^{2}(I^{+}_{xy3}+I^{-}_{xy3})= 0.
\end{equation}
When $\tau\geq \mathcal{O}(1)$, we find that the term $\sum_{x}\sin^2(\omega_{x3}\tau/2)$ is at most a constant. When $\tau$ is integer multiples of $2\pi/\omega_{13}$ ($2\pi/\omega_{23}$), the first (second) term of the sum is zero and the second (first) term is small. So in that case the average error $\langle\epsilon\rangle$ is at least $\mathcal{O}(1/np)$. In fact the error oscillates, but never goes to zero. We bound the maximum of the average error in this scenario. In the case where $\tau<\mathcal{O}(1)$, we can approximate $\sin^2(\tau)\approx \tau^2$. Finally we have that
\begin{equation}
\label{eq:average_error_supp_2}
\langle\epsilon\rangle = 
  \begin{cases} 
   \mathcal{O}\left(\dfrac{\tau}{p\sqrt{n}}\right) & \text{if } \tau <\mathcal{O}(1) \\
      \mathcal{O}\left(\dfrac{1}{p\tau\sqrt{n}}\right)    & \text{if } \tau \geq \mathcal{O}(1)
  \end{cases}.
\end{equation}
$\bullet$ When $p\geq p_{temp}=\log(n)/\sqrt{n}:$ The algorithm is optimal for any $0\leq\tau\leq \mathcal{O}(\sqrt{n})$. 
\\
\\
$\bullet$ When $p_{static}=\log^{3/2}(n)/n\leq p\leq p_{temp}:$ There exists two regimes of temporality, 
$\tau_{\text{slow}}=\mathcal{O}(\log(n)/(p\sqrt{n})$ and $\tau_{\text{fast}}=\mathcal{O}(p\sqrt{n}/\log(n))$ such that if the topology of the network changes faster than $\tau_{\text{fast}}$ or slower than $\tau_{\text{slow}}$, the algorithm is optimal. 
\\
\\
%%%%%%%%%%%%%%%%%%%%%%%%%%%%%%%%%%%%%%%%%%%%%%%%%%%%%%%%%%%%%%%%%%%%
\textbf{II.~Optimality of quantum search on random temporal networks when $p<p_{static}$---} Here we demonstrate that the spatial search algorithm is optimal for random temporal networks even when each of the underlying static networks are below the percolation threshold. We consider the evolution  of the quantum state averaged over all possible realizations of a random graph. The number of possible realizations of $G(n,p)$ is $|G|=2^{N}$, where $N=\binom{n}{2}$. The average dynamics of the algorithm on a random temporal network after one time step $\tau$ is given by the following superoperator
\begin{align}
\label{eq:superoperator_definition}
\Phi(\rho)=\sum_{r=1}^{|G|}p_r e^{-iH_r\tau}\rho e^{iH_r\tau},
\end{align}
where $p_r$ is the probability of the $r$th realization and $H_r=\ket{w}\bra{w}+\gamma A_{G_r}$ with $A_{G_r}$ being the adjacency matrix corresponding to the $r$th realization of $G(n,p)$. Thus the evolution of the algorithm after $m=\sqrt{n}/\tau$ time steps is given by $\Phi^m(\rho)$. Using the Baker-Campbell-Hausdorff formula, the first order expansion of the superoperator yields
\begin{align}
\Phi(\rho)&=\left\langle\rho-i\tau[H_r,\rho]\right\rangle+\delta\\
          &=\rho-i\tau[\ket{w}\bra{w}+\ket{s}\bra{s},\rho]+\delta\\
          &=\Phi_0(\rho)+\delta,
\end{align}
where the second step follows because expectation value of each entry of $A_{G_r}$ is $p$ and so $\langle A_{G_r}\rangle=np\ket{s}\bra{s}$. Thus $\langle H_r\rangle=\ket{w}\bra{w}+\ket{s}\bra{s}$ and the error $\epsilon$ induced by this truncation is the sum of all higher order terms given by
\begin{equation}
\label{eq:totalerrorpertimestep}
\delta\leq\sum_{k=2}^{\infty}\frac{\tau^k}{k!}\left\langle||H^k_r||\right\rangle,
\end{equation} 
where $||X||$ refers to the spectral norm of matrix $X$.
Let us consider the $k$th order term of $\delta$. This is
\begin{align}
\label{eq:kthordererror}
\delta_k&=\frac{\tau^k}{k!}\left\langle||H^k_r||\right\rangle\\
          &\leq \frac{\tau^k}{k!}\left\langle||H_r||^k\right\rangle =  \frac{\tau^k}{k!}\left\langle||\ket{w}\bra{w}+\gamma A_{G_r}||^k\right\rangle\\
          &\leq \frac{\tau^k}{k!}\left\langle||1+\gamma A_{G_r}||^k\right\rangle.
\end{align}
Also notice that the total error after $m\tau$ time steps is 
\begin{equation}
\epsilon=||\Phi^m(\rho)-\Phi^m_0(\rho)||\leq m\delta.
\end{equation}
Bounding the error $\delta$ (and subsequently $\epsilon$) hinges upon bounding $||A_{G_r}||$. We will consider different regimes of $p$ and obtain bounds on $\tau$ for which the $\epsilon=\mathcal{O}(1/\log(n))$.

$\bullet$ When $\log(n)/n\leq p<\log^{3/2}(n)/n :$ The maximum degree of each node ($d_{max}$) of the random graphs are not far from the average degree $np$. In fact it is known that $d_{max}/np=\mathcal{O}(1)$. Also since for any graph $G$, $||A_G||\leq d_{max}$, we have that 
in this regime of $p$, $\delta_k\leq \mathcal{O}(\tau^k/k!)$. This gives us that $\delta\leq \mathcal{O}(\tau^2)$. Subsequently, $\epsilon\leq\mathcal{O}(\tau\sqrt{n})$ which implies that for $\epsilon$ to be bounded, $\tau<1/(\sqrt{n}\log(n))$.
\\
\\
$\bullet$ When $c/n<p\leq \log(n)/n$ such that $c>1:$ The underlying random graphs are no longer connected. The maximum degree of each node is no longer close to the average degree.  In this regime thus we use the trivial bound that $d_{max}\leq n-1\approx n$ and obtain that $\delta\leq \mathcal{O}(\tau^2/p^2)$ and hence $\epsilon\leq \mathcal{O}(\sqrt{n}\tau/p^2)$. In fact $\tau<1/(n^{5/2}\log(n))$ is sufficient for the error to be bounded. Using a better bound for $||A_{G_r}||$ would improve the bound on $\tau$.
\\
\\
$\bullet$ When $p\leq 1/n:$ In this regime $d_{max}<<n-1$ and hence we use a different bound for $||A_{G_r}||$. As there are very few links in the underlying random graphs, we use the fact that for any graph $G$, $||A_{G}||\leq |E(G)|$, where $|E(G)|$ is the total number of edges of graph $G$. For a random graph $G(n,p)$ the probability of having $l$ edges follows a binomial distribution, i.e. $\binom{N}{l}p^{l}(1-p)^{N-l}$. Thus Eq.~\eqref{eq:kthordererror} can be written as 
\begin{align}
\label{eq:kthordererror_2}
\delta_k\leq\frac{\tau^k}{k!}\sum_{l=0}^{N}\binom{N}{l}p^{l}(1-p)^{N-l}(1+\gamma l)^k
\end{align}
As $p\leq 1/n$, we have $\gamma=1/(np)\geq 1$. So
\begin{align}
\label{eq:kthordererror_3}
\delta_k&\leq\frac{{(2\gamma\tau)}^k}{k!}\sum_{l=0}^{N}\binom{N}{l}l^k p^{l}(1-p)^{N-l}=\frac{{(2\gamma\tau)}^k}{k!}B_k(N,p),
\end{align}
where $B_k(N,p)$ represents the $k$th moment of a Binomial distribution. Now there are two distinct case. Firstly, when $p<1/n^2$, $B_k(N,p)=\mathcal{O}(Np)=\mathcal{O}(n^2p/2)$ and secondly for $1/n^2\leq p\leq 1/n$, $B_k(N,p)=\mathcal{O}(N^kp^k)$. We deal with these cases separately. So when $p<1/n^2$ we have
\begin{align}
\label{eq:errorpertimestep_bound}
\delta&\leq\sum_{k=2}^{\infty}\epsilon_k\\
        &\leq \frac{n^2p}{2}\sum_{k=2}^{\infty}\frac{{(2\gamma\tau)}^k}{k!}=\frac{n^2p}{2}(e^{2\gamma\tau}-1-2\gamma\tau)\\
        &\leq \mathcal{O}(2n^2p\gamma^2\tau^2).
\end{align}
 So when $p<1/n^2$ we have that the total error is
\begin{align}
\label{eq:totalerror_case1}
\epsilon&\leq\mathcal{O}\left(m2n^2p\gamma^2\tau^2\right)=\mathcal{O}\left(\frac{\sqrt{n}\tau}{p}\right).    
\end{align}
So the error $\epsilon=\mathcal{O}(1/\log(n))$ as long as $\tau<p/(\sqrt{n}\log(n))$. 
Now for the case where $1/n^2\leq p\leq 1/n$ we have $B_k(N,p)=\mathcal{O}(N^kp^k)$ and a similar derivation yields that $\Delta$ is bounded as long as $1/(n^{5/2}\log(n))$. Thus we find that the dynamics of the algorithm is well approximated by the superoperator $\Phi_0(\rho)$ and the dynamics is restricted to the two dimensional space spanned by the target state $\ket{w}$ and the (almost) initial state $\swbar$. Thus after a time $T=m\tau=\mathcal{O}(\sqrt{n})$, the probability of finding the solution state $\ket{w}$, $P_w(T)\approx 1$.
\bibliographystyle{unsrt}
\bibliography{Bibliography}

\begin{thebibliography}{10}

\bibitem{kempe2000connectivity}
David Kempe, Jon Kleinberg, and Amit Kumar.
\newblock Connectivity and inference problems for temporal networks.
\newblock In {\em Proceedings of the thirty-second annual ACM symposium on
  Theory of computing}, pages 504--513. ACM, 2000.

\bibitem{mucha2010community}
Peter~J Mucha, Thomas Richardson, Kevin Macon, Mason~A Porter, and Jukka-Pekka
  Onnela.
\newblock Community structure in time-dependent, multiscale, and multiplex
  networks.
\newblock {\em science}, 328(5980):876--878, 2010.

\bibitem{holme2012temporal}
Petter Holme and Jari Saram{\"a}ki.
\newblock Temporal networks.
\newblock {\em Physics reports}, 519(3):97--125, 2012.

\bibitem{starnini2012random}
Michele Starnini, Andrea Baronchelli, Alain Barrat, and Romualdo
  Pastor-Satorras.
\newblock Random walks on temporal networks.
\newblock {\em Physical Review E}, 85(5):056115, 2012.

\bibitem{barabasi2016temporal}
Aming Li, Sean~P Cornelius, Yang-Yu Liu, Long Wang, and Albert-L{\'a}szl{\'o}
  Barab{\'a}si.
\newblock The fundamental advantages of temporal networks.
\newblock {\em arXiv preprint arXiv:1607.06168}, 2016.

\bibitem{Childs_spatial_search}
Andrew~M. Childs and Jeffrey Goldstone.
\newblock Spatial search by quantum walk.
\newblock {\em Physical Review A}, 70:022314, 2004.

\bibitem{Meyer_symmetry}
Jonatan Janmark, David~A. Meyer, and Thomas~G. Wong.
\newblock Global symmetry is unnecessary for fast quantum search.
\newblock {\em Physical Review Letters}, 112:210502, 2014.

\bibitem{meyer2}
David~A. Meyer and Thomas~G. Wong.
\newblock Connectivity is a poor indicator of fast quantum search.
\newblock {\em Physical Review Letters}, 114:110503, Mar 2015.

\bibitem{dimred}
Leonardo Novo, Shantanav Chakraborty, Masoud Mohseni, Hartmut Neven, and Yasser
  Omar.
\newblock Systematic dimensionality reduction for quantum walks: Optimal
  spatial search and transport on non-regular graphs.
\newblock {\em Scientific Reports}, 5:13304, 2015.

\bibitem{li2016continuous}
Shanshan Li and Stefan Boettcher.
\newblock Continuous time quantum walks in finite dimensions.
\newblock {\em arXiv preprint arXiv:1607.05317}, 2016.

\bibitem{Farhi_analog_grover}
Edward Farhi and Sam Gutmann.
\newblock Analog analogue of a digital quantum computation.
\newblock {\em Physical Review A}, 57:2403, 1998.

\bibitem{ER59}
Paul Erd{\H{o}}s and Alfred R{\'{e}}nyi.
\newblock On random graphs. {I}.
\newblock {\em Publ. Math. Debrecen}, 6:290--297, 1959.

\bibitem{ER60}
Paul Erd\H{o}s and Alfred R\'{e}nyi.
\newblock {On the evolution of random graphs}.
\newblock {\em Publications of the Mathematical Institute of the Hungarian
  Academy of Sciences}, 5:17--61, 1960.

\bibitem{almost_all}
Shantanav Chakraborty, Leonardo Novo, Andris Ambainis, and Yasser Omar.
\newblock Spatial search by quantum walk is optimal for almost all graphs.
\newblock {\em Physical Review Letters}, 116:100501, Mar 2016.

\bibitem{QCN2}
Mauro Faccin, Tomi Johnson, Jacob Biamonte, Sabre Kais, and Piotr Migda\l{}.
\newblock Degree distribution in quantum walks on complex networks.
\newblock {\em Physical Review X}, 3:041007, Oct 2013.

\bibitem{QCN1}
Mauro Faccin, Piotr Migda\l{}, Tomi~H. Johnson, Ville Bergholm, and Jacob~D.
  Biamonte.
\newblock Community detection in quantum complex networks.
\newblock {\em Physical Review X}, 4:041012, Oct 2014.

\bibitem{leung2010coined}
Godfrey Leung, Paul Knott, Joe Bailey, and Viv Kendon.
\newblock Coined quantum walks on percolation graphs.
\newblock {\em New Journal of Physics}, 12(12):123018, 2010.

\bibitem{kollar2012coined}
B{\'a}lint Koll{\'a}r, Tam{\'a}s Kiss, Jaroslav Novotn{\`y}, and Igor Jex.
\newblock Asymptotic dynamics of coined quantum walks on percolation graphs.
\newblock {\em Physical Review Letters}, 108(23):230505, 2012.

\bibitem{darazs2013continuous}
Zolt{\'a}n Dar{\'a}zs and Tam{\'a}s Kiss.
\newblock Time evolution of continuous-time quantum walks on dynamical
  percolation graphs.
\newblock {\em Journal of Physics A: Mathematical and Theoretical},
  46(37):375305, 2013.

\bibitem{cerfroland}
J\'er\'emie Roland and Nicolas~J. Cerf.
\newblock Noise resistance of adiabatic quantum computation using random matrix
  theory.
\newblock {\em Phys. Rev. A}, 71:032330, Mar 2005.

\bibitem{Note1}
The search Hamiltonians $H_{search}=\mathinner {|{w}\delimiter "526930B
  }\mathinner {\delimiter "426830A {w}|}+\gamma A_G$ and
  $H_{search}=-\mathinner {|{w}\delimiter "526930B }\mathinner {\delimiter
  "426830A {w}|}-\gamma A_G$ are equivalent and leave the outcome of the
  algorithm unaltered. We shall use them interchangeably.

\bibitem{Note2}
In our analysis when $p\geq \protect \qopname \relax o{log}^{3/2}(n)/n$, we
  neglect terms of $\protect \mathcal {O}(1/(np))$ and we fix $\mathinner
  {\delimiter "426830A {\epsilon }\delimiter "526930B }=1/\protect \qopname
  \relax o{log}(n)$ to be the upper bound of error throughout and obtain bounds
  for $\tau $ and $p$.

\bibitem{furedi1981eigenvalues}
Zolt{\'a}n F{\"u}redi and J{\'a}nos Koml{\'o}s.
\newblock The eigenvalues of random symmetric matrices.
\newblock {\em Combinatorica}, 1(3):233--241, 1981.

\bibitem{vu2007}
Van~H. Vu.
\newblock Spectral norm of random matrices.
\newblock {\em Combinatorica}, 27(6):721--736, 2007.

\bibitem{barabasi_review}
R{\'e}ka Albert and Albert-L{\'a}szl{\'o} Barab{\'a}si.
\newblock Statistical mechanics of complex networks.
\newblock {\em Reviews of Modern Physics}, 74(1):47, 2002.

\bibitem{hein2009wave}
Birgit Hein and Gregor Tanner.
\newblock Wave communication across regular lattices.
\newblock {\em Physical Review Letters}, 103(26):260501, 2009.

\bibitem{tanner_graphene}
Iain Foulger, Sven Gnutzmann, and Gregor Tanner.
\newblock Quantum search on graphene lattices.
\newblock {\em Physical Review Letters}, 112(7):070504, 2014.

\bibitem{scholtes2014causality}
Ingo Scholtes, Nicolas Wider, Ren{\'e} Pfitzner, Antonios Garas, Claudio~J
  Tessone, and Frank Schweitzer.
\newblock Causality-driven slow-down and speed-up of diffusion in non-markovian
  temporal networks.
\newblock {\em Nature communications}, 5, 2014.

\end{thebibliography}

%\bibliographystyle{unsrt}
%\bibliography{biblio2}
\end{document}